\documentclass[11pt]{article}
\usepackage{geometry}                
\usepackage{graphicx}
\usepackage{amssymb}
\usepackage{comment,color}
\usepackage{epstopdf}
\usepackage{hyperref}
\title{Double pulses and cascades above 2 PeV in IceCube}
\author{Andrea Palladino\footnote{Gran Sasso Science Institute, andrea.palladino@gssi.infn.it}, Giulia Pagliaroli\footnote{LNGS}, Francesco L. Villante\footnote{Universit\'a degli studi dell'Aquila, LNGS}, Francesco Vissani\footnote{LNGS, Gran Sasso Science Institute}}

\graphicspath{{figures/}}
\begin{document}
\maketitle
\begin{abstract}
IceCube collaboration has seen an unexpected population of high energy
neutrinos compatible with an astrophysical origin. We
consider two categories of events that can help to diagnose cosmic neutrinos:
double pulse, that may allow us to clearly discriminate the
cosmic component of $\nu_\tau$; cascades with deposited energy above 2 PeV,
including events produced by $\overline{\nu}_e$ at the Glashow resonance,
that can be used to investigate the neutrino production
mechanisms. 
We show that one half of the double pulse signal 
is due to the neutrinos spectral region already probed
by IceCube. By normalizing to HESE data, we find that 10 more years
are required to obtain 90$\%$ probability to observe a double pulse. 
The cascades above 2 PeV provide us a sensitive probe of the high
energy tail of the neutrino spectrum and are potentially observable, 
but even in this case, the dependence on type of the source is mild.
In fact we find that $pp$ or $p\gamma$ mechanisms give a difference
in the number of cascades above 2 PeV of about 25 $\%$ that can be
discriminated at $2\sigma$ in $\sim50$ years of data taking.
\end{abstract}
\maketitle

\section{Introduction}
In four years of data taking, IceCube has observed \textcolor{black}{$32$} High Energy
Starting Events (HESE) with deposited energies between 60 TeV and 2 PeV
\cite{Aartsen:2013jdh,Aartsen:2014gkd,Aartsen:2015knd,IPA}.
The scientific debate about the origin of these events is extremely
lively. There is little doubt that cosmic neutrinos have been seen,
but their origin is not yet understood.

 In this work, we focus our attention on two specific classes of events,
not yet observed, that can give us precious information
on the extra terrestrial component of neutrinos flux: the so-called double
pulse events, due to tau-neutrinos \cite{Aartsen:2015tau},
and the cascades above 2 PeV that include events due to electron
antineutrinos interacting at the Glashow resonance.

 The $\nu_\tau$ are not expected to be produced in astrophysical sources
(nor in the atmosphere)
but they are predicted to be a non negligible component of the cosmic
neutrino flux due to flavor oscillations
\cite{beacom1,Vissani:2013iga,Palladino:2015zua,flv1,mena,flv2,flv3,Pagliaroli}
and thus represent a distinctive signature of a cosmic
population.
At low energy,  it is impossible to distinguish cascades produced by
charged current (CC) interactions of $\nu_\tau$
from those produced by CC-interactions of $\nu_e$ and neutral current (NC)
interactions
of all neutrino flavors. The only way to tag $\nu_\tau$ is to observe a
double pulse in the detector \cite{Learned:1994wg,Cowen:2007ny, Abbasi:2012cu},
which is produced by the CC-interaction of $\nu_\tau$, when $\tau$ is
produced, followed by a second energy release, when
the $\tau$ decays \footnote{We use the terminology ”double pulse”, recently introduced by the IceCube collaboration \cite{Aartsen:2015tau}, rather than with the traditional terminology ”double bang” \cite{Learned:1994wg}, in order to emphasize that we adopt the same experimental requirements of the IceCube collaboration. It is possible that in future years, experimental cuts will be optimised further, with a possible increased number of events.}. A very recent analyses from IceCube
\cite{Aartsen:2015tau}, dedicated to the search of these events
with different topology with respect to tracks and cascades, reported a null
result. We discuss
the implications of this result and the perspective for future $\nu_\tau$
detection.

The second class of events considered in this paper are cascades with
deposited energy above 2 PeV; these events
can be produced by deep inelastic scattering (DIS) of high energy $\nu_e$
and $\nu_\tau$
and by $\overline{\nu}_e$ interacting with electrons through
Glashow resonance \cite{Gres}. As already discussed in
\cite{Berezinsky,Anchordoqui:2004eb,Bhattacharya:2011qu,Pakvasa,flv2},
the rate of these events depends on the neutrino production mechanisms.
In particular, since Glashow resonance is only possible for
$\overline{\nu}_e$,
a larger signal is expected if neutrinos are produced by $pp$ collisions
with respect to
the case of $p \gamma$ interactions (see \cite{aha1,aha2,Villante:2008qg} for a review on the spectra of secondary particles produced in $pp$ and $p\gamma$ interaction), being indeed the
antineutrino fraction
larger in the first case. The possibility to discriminate among the two
mechanisms depends on
the relative contributions of events produced by DIS and Glashow resonance.
We perform a realistic calculation of these contributions. Differently
from previous work on the subject \cite{Pakvasa}, we
discuss the role of leptonic channels in Glashow resonance that can be
correctly evaluated only
if the difference between the incoming neutrino energy and the energy
deposited in the detector is taken into account.

 The expected rates of both classes of events depend on the assumed
neutrino energy distribution.
Our nominal hypothesis is that the cosmic neutrino spectra are described
by single power law that extends until 10 PeV.
We consider the neutrino spectral index as a free parameter and we fix the
normalization of fluxes
by requiring that they produce the events observed by IceCube at low
energies (i.e. below 2 PeV).
We thus obtain the expected rates of double pulse and cascades above 2 PeV
as a function of the
slope of the neutrino spectrum. This permits us to discuss the relevance of
the assumed
neutrino energy distribution for future $\nu_\tau$ detection, for the
discrimination of $pp$ or
$p \gamma$ production mechanism and/or for the observation of high energy
cutoff, automatically
implementing the present information provided by IceCube at low energy.

The plan of the paper is the following: in Sec.\ref{sec1} we describe 
our assumptions on the cosmic neutrino flux, in Sec.\ref{sec2} we
calculate the expected number of double pulse events in IceCube and in
Sec.\ref{sec3} the expected number of cascades with deposited energy
avove 2 PeV. In Sec.\ref{sec4} we made a comparison between our
results and previous works on these subjects and finally, in Sec.\ref{sec5}, we draw our
conclusions.

\section{The cosmic neutrino flux}
\label{sec1}

We assume that the total flux of cosmic neutrinos (and antineutrinos) has
an isotropic distribution
and that the spectrum can be described by a power law
\begin{equation}
\frac{d \phi_{\nu+\overline{\nu}}}{dE_\nu} = F(\alpha) \
\frac{1}{\mbox{PeV} \ \mbox{m}^2 \ \mbox{year}}
\bigg(\frac{E_\nu}{\mbox{PeV}}\bigg)^{-\alpha}
\label{PL}
\end{equation}
that extends till $E_{\rm cut}=10 \,{\rm PeV}$. Recalling that neutrinos take about 1/20 of the energy of the parent proton in cosmic ray  interactions, this means that we are considering protons with energies up to 200 PeV in
their sources.
It is generally expected that, due to flavor oscillations, a cosmic
neutrino population is characterized
by a flavor content $(1/3 : 1/3 : 1/3)$ independently on the specific
production mechanism.
In reality, a certain imprint of the neutrino production mechanism does
remain, as it is discussed e.g. in \cite{Palladino:2015zua,Vissani:2013iga,flv3}.
The fluxes divided per flavors can be generally given as:
\begin{eqnarray}
\nonumber
\frac{d \phi_{\nu_e+\overline{\nu}_e}}{dE_\nu} &=&
\left(\frac{1}{3}+\frac{2 P_1}{3} \right)  \frac{d
\phi_{\nu+\overline{\nu}}}{dE_\nu}  \\
\nonumber
\frac{d \phi_{\nu_\mu+\overline{\nu}_\mu}}{dE_\nu} &=&
\left(\frac{1}{3}-\frac{P_1}{3} +\frac{2 P_2}{3}\right) \frac{d
\phi_{\nu+\overline{\nu}}}{dE_\nu}\\
\frac{d \phi_{\nu_\tau+\overline{\nu}_\tau}}{dE_\nu} &=&
\left(\frac{1}{3}-\frac{P_1}{3} -\frac{2 P_2}{3} \right) \frac{d
\phi_{\nu+\overline{\nu}}}{dE_\nu}
\end{eqnarray}
where $P_1$ and $P_2$ are (small) parameters described in \cite{flv2} that
are determined
by the neutrino flavor content at the source (i.e. before oscillations).
In the following, we consider the case of neutrino
produced by charged pion decays for which $P_1=0.000 \pm 0.029$,
$P_2=0.010 \pm 0.007$;
the errors are obtained by propagating uncertainties in neutrino
oscillation parameters.

The normalization of the flux $F(\alpha)$ is obtained by requiring that
the number of events, due to cosmic neutrinos,
reproduces the results obtained by IceCube at low energies (i.e. between 60
Tev and 2 PeV).
In three years of data taking, IceCube has observed $N_{\rm tot} =20$ events
against an expected background
of $N_{\rm B} = 2.8$  events from atmospheric muons and
neutrinos\footnote{We assume
that the prompt atmospheric neutrinos give negligible
contributions, as it is required by the arrival angles
distributions of IceCube events \cite{Aartsen:2014gkd}. Anyway it will
be important to measure also this component of atmospheric neutrinos
in the future. \textcolor{black}{We normalize the neutrino spectrum by
using the three year IceCube results because the complete
information for this data set is provided in Supp.Tab. IV of \cite{Aartsen:2014gkd} 
allowing
us to crosscheck and validate our conclusions.}}. We require that the number of events
from astrophysical neutrinos,
calculated as:
\begin{equation}
N=T \int^{2 \ \rm PeV} dE_\nu \; \sum_{\ell=e,\mu,\tau} A_{\ell}(E_\nu)
\, \frac{d \phi_{\nu_\ell +\overline{\nu}_\ell}}{dE_\nu}
\end{equation}
where $T$ is the observation time and $A_{\ell}(E_\nu) $ are the {\em
effective areas} for the various neutrino
flavors given in \cite{Aartsen:2013jdh}, is equal to $N = N_{\rm tot}-N_{\rm B}$.
We introduced an upper integration limit to mimic
the effect of the IceCube observation threshold at 2 PeV.

By following the above procedure, we obtain the flux normalization:
\begin{equation}
F(\alpha)=0.12 \cdot [0.95-0.9(\alpha-2)]
\label{fa}
\end{equation}
Note that the coefficient $F(\alpha)$ determines the flux of cosmic
neutrinos at 1 PeV.
In the power law assumption, see Eq.(\ref{PL}), this quantity is relatively
well constrained being
equal to $\sim 0.11$ for $\alpha=2$ and
$\sim 0.05$ for
$\alpha=2.6$.

\section{Tau neutrinos and double pulse events}
\label{sec2}

As stated in the Introduction, one of the goals of this work is to
discuss the detection of the $\nu_\tau$ component of the high energy (HE) neutrino
flux providing the proof of existence of a cosmic population. 
We are interested to investigate the dependence of the expected number
of double pulse events from the energy distribution of cosmic neutrinos and from the IceCube observation time.  

In order to perform this calculation we need the effective area for
double pulse events, $A^{\mbox{\tiny 2P}}_\tau$, recently published
by the IceCube collaboration \cite{Aartsen:2015tau}. Following the
IceCube prescription, 
the expected number of double pulse events in the observation time $T$ is
\begin{equation}
N_{\mbox{\tiny 2P}}(\alpha) =T\times  \int^{E_{\mbox{\tiny cut}}} \ dE_{\nu} \ A^{\mbox{\tiny 2P}}_\tau(E_\nu)
 \frac{d\phi_{\nu_\tau+\overline{\nu}_\tau}}{dE_\nu}  
\label{tau1}
\end{equation}
where 
the differential flux of the $\nu_\tau$ component,
$\frac{d\phi_{\nu_\tau+\overline{\nu}_\tau}}{dE_\nu}$ is normalized 
to reproduce the HESE events observed by IceCube (see previous section). 

In order to discuss the dependence of $N_{\mbox{\tiny 2P}}$ from the
spectral index $\alpha$, it is useful to give 
an analytical description of the IceCube effective area. Considering
that double pulse events are a {\em subset} of the events caused by CC
tau neutrino interactions, we describe the effective area as,
\begin{equation}
A^{\mbox{\tiny 2P}}_\tau(E_\nu) = \epsilon_{\mbox{\tiny 2P}}\times\eta_{\mbox{\tiny CC}} \times A_{\tau}(E_\nu) \times  P_{\mbox{\tiny 2P}}(E_\nu,L_{\mbox{\tiny min}}) 
\label{2p}
\end{equation}
where $A_{\tau}(E_\nu)\approx 13.4 \mbox{ m}^2 (E_\nu/\mbox{PeV})^{0.455}$ is the effective area for $\nu_\tau$
calculated in \cite{Aartsen:2013jdh}, the factor $\eta_{\mbox{\tiny    CC}} = (1+{\sigma_{\mbox{\tiny NC}}}/{\sigma_{\mbox{\tiny      CC}}})^{-1}\approx 0.7$ gives the fraction of $\nu_\tau$ interactions that are due to CC processes
and the constant $ \epsilon_{\mbox{\tiny 2P}}<1$ describes the
effect of geometrical and quality cuts implemented by IceCube for the
search of these events.
%
%
The function $P_{\mbox{\tiny 2P}}(E_\nu,L_{\mbox{\tiny min}})$
describes the probability that a neutrino with energy $E_{\nu }$
produces a tau traveling more than $L_{\mbox{\tiny min}}$ before it
decays, where $L_{\mbox{\tiny min}}$ is the minimum distance to give rise
to an observable double pulse in the detector. We expect that
$L_{\mbox{\tiny min}}$ is of the order of tens of meters\footnote{One
  can implement a condition  for containment replacing 
$P_{\mbox{\tiny 2P}}(E_\nu,L_{\mbox{\tiny min}})\to P_{\mbox{\tiny 2P}}(E_\nu,L_{\mbox{\tiny
    min}})-P_{\mbox{\tiny 2P}}(E_\nu,L_{\mbox{\tiny max}})$ with $L_{\mbox{\tiny
    max}}\sim 0.5$ km; 
we checked that the changes are not conspicuous in the range of energies of interest.}, that is the typical distance between the DOMs \cite{Aartsen:2015tau}.

The taus produced in CC-DIS have an average energy equal to:
\begin{equation}
E_{\tau}= (1-\langle y \rangle) E_{\nu} \simeq \frac{3}{4}
E_{\nu}
\label{Etau}
\end{equation}
where $\langle y \rangle$ is the mean inelasticity which is nearly
constant in the energy range 
that we are considering \cite{Gandhi}. If we neglect $\tau$ energy
dispersion and assume the one-to-one relationship between $E_\tau$ and
$E_\nu$ expressed by Eq.(\ref{Etau}), the probability $P_{\mbox{\tiny 2P}}(E_\nu,L_{\mbox{\tiny min}})$ is given by:
\begin{equation}
P_{\mbox{\tiny 2P}}(E_\nu,L_{\mbox{\tiny min}})= \mbox{exp} \bigg[-\frac{E_{\mbox{\tiny min}}(L_{\mbox{\tiny min}})}{E_\nu} \bigg]
\end{equation}
where $E_{\mbox{\tiny min}}$ represents the minimum neutrino energy
which is necessary to produce a tau with decay length
larger than $L_{\mbox{\tiny min}}$. This can be calculated as:
\begin{equation}
E_{\mbox{\tiny min}}= \frac{ L_{\mbox{\tiny min}}}{c \ t_\tau}\times \frac{m_\tau  c^2 }{1-\langle y \rangle} = 3.3 \mbox{ PeV} \bigg(\frac{ L_{\mbox{\tiny min}}}{120  \ \mbox{m}}\bigg)
\end{equation}
with $m_{\tau} c^2=1.777 \mbox{ GeV}$ and $t_\tau = 0.29 \cdot10^{-12} \mbox{ s}$.

\begin{figure}[t]
\centering
\includegraphics[scale=0.7]{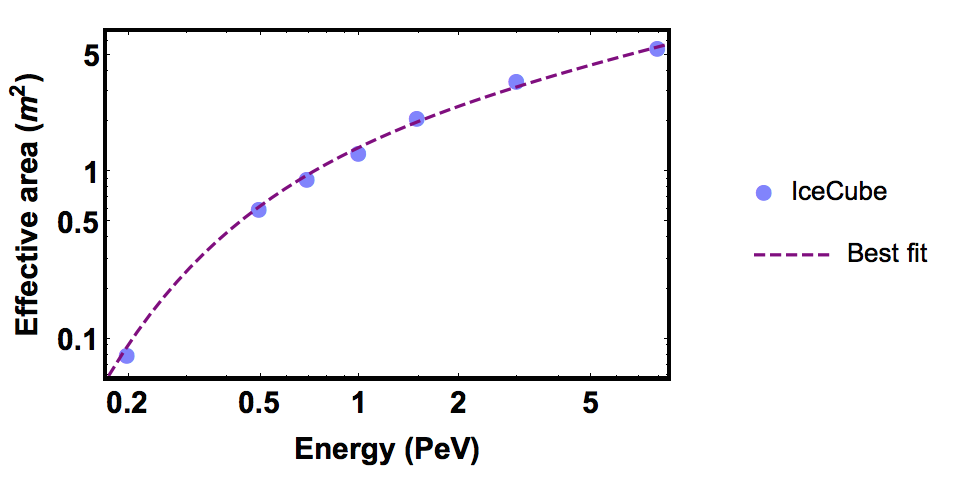}
\caption{\em\small Effective areas of double pulse. The points are the values given by IceCube \cite{Aartsen:2015tau} while the line is the parameterization described in the text.}
\label{2bangeff}
\end{figure}

Using the expression in Eq.(\ref{2p}), we find that the effective area of IceCube, in the energy region from 0.1 to 10 PeV, is reasonably well described setting 
\begin{equation}
\epsilon_{\mbox{\tiny 2P}}=0.25\mbox{ and }   E_{\mbox{\tiny min}}=0.5  \mbox{ PeV}
\end{equation} 
that corresponds to $L_{\mbox{\tiny min}}=18$ m. In other words the following parameterized expression for the effective area can be used:
\begin{equation}
A^{\mbox{\tiny 2P}}_\tau= \bar{A}_{\mbox{\tiny 2P}}\times
  \bigg(\frac{E_\nu}{\mbox{PeV}} \bigg)^{\!\beta} \ \mbox{exp}\bigg(-\frac{E_{\mbox{\tiny min}}}{E_{\nu}} \bigg) 
  \mbox{ with } \left\{\begin{array}{l}\bar{A}_{\mbox{\tiny 2P}}=2.33 \mbox{ m}^2 \\[1ex] \beta=0.455\\[1ex]  E_{\mbox{\tiny min}}=0.5\mbox{ PeV}\end{array} \right\},
\end{equation}
as showed in Fig.\ref{2bangeff} for a direct comparison of the IceCube effective area.

Using the previous expression it is possible to obtain an accurate analytical formula for the expected
number of double pulse events
\begin{equation}
N_{\mbox{\tiny 2P}}(\alpha) = \frac{F(\alpha)}{3} \times
 \frac{\bar{A}_{\mbox{\tiny 2P}}}{\mbox{m}^2}\times 
\frac{T}{\mbox{yr}}\times  
  \left( \frac{E_{\mbox{\tiny min}}}{\mbox{PeV}}
  \right)^{\beta-\alpha+1} \Gamma\left(\alpha-\beta-1,
    \frac{E_{\mbox{\tiny min}}}{E_{\mbox{\tiny cut}}}\right)
\end{equation}
where $\Gamma$ is the incomplete gamma function and the normalization of
$\nu_\tau + \overline{\nu}_\tau$ flux is assumed to be $F(\alpha)/3$,
as it is expected for neutrinos produced by charged pions decays with
few $\%$ uncertainty due to errors in the neutrino oscillation
parameters (see previous section for details). To check our result, 
we compare with IceCube calculations in \cite{Aartsen:2015tau} 
finding agreement at the level of few percents.

The above expression allows us to investigate the dependences of the
expected number of double pulse events on the spectral index $\alpha$ and
on the high energy cutoff  $E_{\mbox{\tiny cut}}$ of the neutrino
spectrum. 
In particular, it permits us to
show  that our knowledge of the neutrino spectrum is already
sufficient to make significative predictions. 

The number of double pulse events expected in 4 years of data taking is 0.66,
0.53, 0.41, 0.31 for $\alpha=2.0,2.2,2.4$ and $2.6$, so it is not
surprising that IceCube have not seen double pulse events so far.
Our calculations are done by adopting the nominal cutoff energy 
$E_{\mbox{\tiny  cut}}=10$ PeV. However, the predicted values are not
strongly dependent on the assumed high energy cutoff. For $\alpha =2.0$, the counting rate varies indeed by only
$\sim 25\%$ when the cutoff energy
is varied within the decade $E_{\mbox{\tiny  cut}}=5-50$ PeV. For larger values of
$\alpha$,  the dependence of $N_{\mbox{\tiny 2P}}(\alpha)$  on
$E_{\mbox{\tiny cut}}$ is considerably weaker. 

The dependence of $N_{\mbox{\tiny 2P}}(\alpha)$ on the spectral index
mainly arises from the normalization $F(\alpha)$ of the cosmic
neutrino flux at $1\;{\rm PeV}$, see Eq.(\ref{PL}) . 
The residual dependence on $\alpha$ is relatively weak and affects 
the final results at the few $\%$ level when $\alpha=2.0-2.6$. 
A good approximation for the predicted number of double pulses is
thus given by:
\begin{equation}
N_{\mbox{\tiny 2P}}(\alpha) \approx 1.45 \times \frac{T}{\mbox{year}} \times F(\alpha)
\label{prox1}
\end{equation} 
We remind that the normalization $F(\alpha)$ is constrained within a factor
of $2$ for $2.0 \le \alpha \le 2.6$. \textcolor{black}{As remarked in \cite{Aartsen:2015tau}, the optimal neutrino energy window to see the double pulse events is between 0.1 to 10 PeV.} \textcolor{black}{It is important to remark the consequence of this fact: 
\begin{quote} assuming cosmic origin, a large 
fraction of the double pulse events are generated by a parent neutrino spectrum
which is already observed by IceCube; 
\end{quote}
conversely, a lack of observation would have dramatic implications, either on the origin of these events or on the nature of neutrino oscillations.}
This can be
better appreciated from Fig.\ref{parental}, where we show with a
yellow line the integrand $dN_{\mbox{\tiny 2P}}/dE_{\nu}$ of \textcolor{black}{Eq.(\ref{tau1})}
calculated for $\alpha=2.3$.The function $dN_{\mbox{\tiny  2P}}/dE_{\nu}$ is peaked around 0.5 PeV and approximately one half of
the double pulse signal is due to neutrinos with initial energy below
2 PeV, i.e. to the energy region already probed by
HESE observations in IceCube.

\begin{figure}[t]
\centering
\includegraphics[scale=0.6]{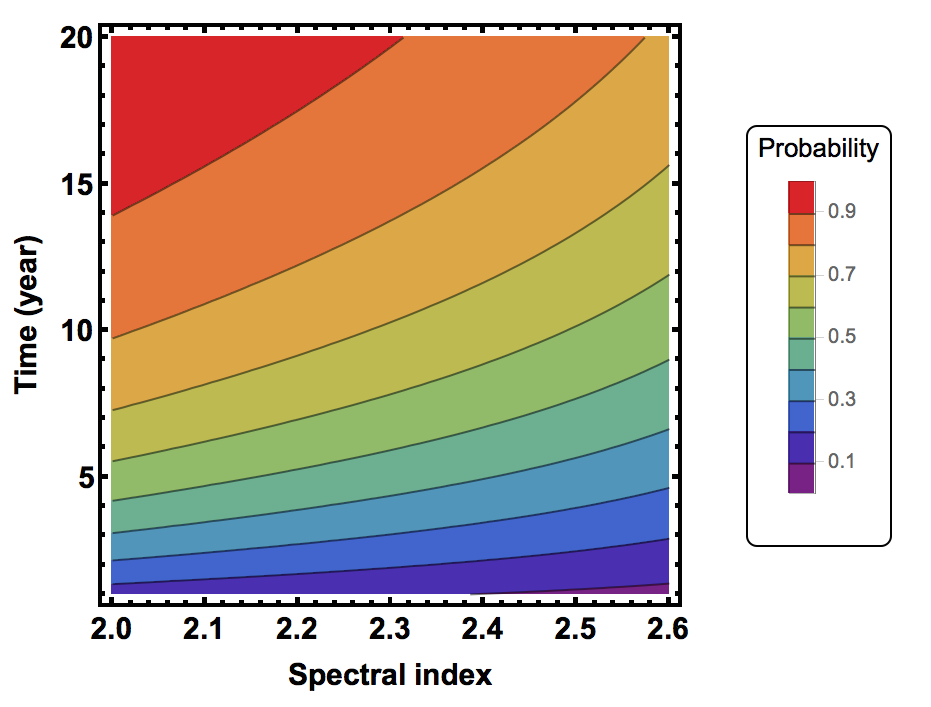}
\caption{\em \small Probability to observe at least one double pulse event as a function of spectral index and number of years.}
\label{2bang}
\end{figure}

Finally, we show in Figure \ref{2bang} the probability to observe at least one double pulse as a function of spectral index and number of years. 
To observe a double pulse in IceCube with a probability greater than
90 $\%$ we must wait about 10 years in the most favorable case
($\alpha=2$), about 15 years in the case $\alpha=2.3$ and
much more if the spectral index is close to $\alpha=2.6$.


\section{Cascades events above 2 PeV}
\label{sec3}

In this section we estimate the number of cascade events with deposited
energy above $2\;{\rm PeV}$  in IceCube.
Two types of cascade events are described: those from deep inelastic scattering (DIS) and 
those produced by the Glashow resonance \cite{Gres}.

\subsection{Cascades from DIS}
\label{disec}
Cascades from DIS are mostly given by CC-interactions of $\nu_e$ and $\nu_\tau$ with a negligible contribution from NC 
interactions of neutrinos of all flavors, as discussed in the following. 
The expected number of cascades from DIS, \textcolor{black}{with deposited energy above 2 PeV}, is given by:
\begin{equation}
N_{\mbox{\tiny DIS}}(\alpha) =  \eta_{\rm CC}  \, T \int  d E_\nu \left[ \frac{d \phi_{\nu_e +
    \overline{\nu}_e}}{d E_\nu}  A^{\mbox{\tiny DIS}}_e(E_\nu) \,
P_e(E_\nu,E_{\rm th}) + \frac{d \phi_{\nu_\tau + \overline{\nu}_\tau}}{d
  E_\nu} A^{\mbox{\tiny DIS}}_\tau(E_\nu)  \, P_\tau (E_\nu ,E_{\rm th})
\right]
\label{DIS}
\end{equation}
where:\\
- $A^{\mbox{\tiny DIS}}_e$ and $A^{\mbox{\tiny DIS}}_\tau$ are the effective area for DIS of
$\nu_e$ and $\nu_\tau$, which are calculated in sect. \ref{disentangling};\\
- the factor $\eta_{\mbox{\tiny CC}}$ is given in the previous section; \\
- the function $P_\ell(E_\nu,E_{\rm th})$ represents the probability
that a CC-DIS event produced by neutrino $\nu_\ell$ of energy
$E_{\nu}$ has a visible energy above $E_{\rm  th} = 2\,{\rm PeV}$. 

\textcolor{black}{In CC-interactions of $\nu_e$ an electromagnetic cascade is produced and the incoming neutrino energy is entirely deposited in the detector, i.e. $E_{\rm dep}=E_\nu$. By using the direct relationship between $E_{\rm dep}$ and $E_\nu$ we can write the probability to observe an event with $E_{\rm dep} \geq E_{\rm th}$ as:
\begin{equation}
P_{e}(E_{\nu},E_{\rm th}) =
\frac{1}{2} \bigg[ 1+\mbox{Erf}\bigg(\frac{E_{\nu}-E_{\rm
    th}}{\sqrt{2} \ E_{\nu} \ \delta  }\bigg)\bigg]
\label{Pe}
\end{equation}  
where `Erf' indicates the error function, $\delta =12 \%$ and we
assumed that the energy resolution for the deposited energy is
described by a Gaussian with a variance $\Delta E_{\rm dep}=\delta
\cdot E_{\rm dep}  $ \cite{Aartsen:2014gkd}. }

\textcolor{black}{In CC-interactions of $\nu_\tau$, a small fraction of the incoming
neutrino energy is carried away by the invisible outgoing neutrinos
produced in $\tau$ decay. If we neglect the energy dispersion of outgoing neutrinos, we can take this into account by writing $E_{\rm dep}=
\eta_{\nu_\tau} E_{\nu}$, where $\eta_{\nu_\tau}=0.8$ is the average energy fraction deposited in the detector by hadrons and charged leptons (see \cite{Vissani:2013iga}). With this assumption the probability $P_{\tau}(E_{\nu},E_{\rm th})$ is obtained from Eq.(\ref{Pe}) by replacing $E_{\nu} \rightarrow \eta_{\nu_\tau} E_{\nu}$. In the above estimate, we neglect that the 17.4$\%$ of taus decays into muon producing track events and this corresponds to overestimating the total number of cascades due to $\nu_e$ and $\nu_\tau$ by 7$\%$ at most.}

\textcolor{black}{In NC-interactions only a small fraction of the initial neutrino
energy is deposited in the detector: on average $E_{\rm dep}=\frac{1}{4} E_\nu$. Therefore only neutrinos of relatively high energy give a contribution to the signal; with the threshold of 2 PeV we need neutrinos with energy around $E_{\nu}=8$ PeV.  We estimated that the contribution
of NC to the total number of events above 2 PeV is equal to few $\%$
when $\alpha=2$ and decreases with the increasing of the spectral index. For this reason we neglected it in the calculation.}
\subsection{Cascades from Glashow resonance}
\begin{table}[t]
\centering
\caption{\em \small Branching ratio of $W^-$ decay} 
\label{brw}
\begin{tabular}{|c|c|}
\hline
 & Branching ratio  \\ 
\hline
$\Gamma (\ell \  \nu) / \Gamma_{total}$  & 10.86 $\pm$ 0.09 $\%$ \\
\hline
$\Gamma (hadrons) / \Gamma_{total}$ & 67.41 $\pm$ 0.27 $\%$  \\
\hline
\end{tabular}
\end{table}
The CC-interaction process $\overline{\nu}_e+e^{-}$, mediated by an
intermediate $W$ boson, has a resonant character at: 
\begin{equation} 
E_{\rm G}=\frac{M_W^2}{2 m_e}=6.32\mbox{ PeV}
\end{equation}
The cross section at $E\simeq E_{\rm G}$ is about 2 order of magnitude
larger than that of DIS  and provides the dominant contribution to the
$\overline{\nu}_e$ interaction rate at few PeVs.

The properties of events produced by Glashow resonance depend
on the final state of the interaction process, i.e. on the $W^{-}$
decay mode. We thus consider separately the different contributions
to the total events number $N_{\rm G}(\alpha)$, \textcolor{black}{with deposited energy above 2 PeV}, obtaining:
\begin{equation}
N_{\rm G}(\alpha) =  T \int  d E_\nu  \,
\frac{d\phi_{\nu_e+\overline{\nu}_e}}{d E_\nu}  \,  A^{\rm G}_e(E_\nu) \,
\frac {\xi_{\bar{\nu}_e}}{\tilde{\xi}_{\bar{\nu}_e}} \,
\left[B_{\rm H} \, P_{\rm H}(E_\nu,E_{\rm th}) + 
\sum_{\ell=e,\tau} B_{\nu\ell} \, P_{\nu\ell}(E_\nu,E_{\rm th})
\right]
\label{GR}
\end{equation}
where:\\
- $A^{\rm G}_e(E_\nu)$ is the effective area for Glashow resonance 
which is calculated in sect. \ref{disentangling};\\
- the parameter $\xi_{\bar{\nu}_e}$ is the fraction of $\overline{\nu}_{e}$ in the electron
neutrino+antineutrino flux. We take as reference the value
$\tilde{\xi}_{\bar{\nu}_e}=1/2$ that is used by IceCube in effective
areas calculations \cite{Aartsen:2013jdh};\\
- the factors $B_{\rm H}$ and $B_{\nu\ell}$ are the branching ratios of
$W^{-}\to{\rm hadrons}$ and $W^{-}\to{\overline \nu}_\ell + \ell$ with
$\ell=e,\, \tau$ respectively, which are given in Tab. \ref{brw}. Note
that we do not include the contribution from $W^{-}\to
\overline{\nu}_\mu + \mu$ because muons produce tracks (not cascades) in the detector;\\
- the functions $P_{\rm H}(E_\nu,E_{\rm th})$ and $P_{\nu\ell}(E_\nu,E_{\rm th})$ 
represent the probability
that an event produced by $\overline{\nu}_e$ of energy $E_{\nu}$ 
through hadronic or leptonic decay modes has a deposited energy above 
$E_{\rm  th} = 2\,{\rm PeV}$. 

When $W^-$ decays in hadrons, an hadronic shower is produced and 
all the energy of the incoming $\overline{\nu}_e$ 
is deposited in the detector, i.e. $E_{\rm dep} = E_\nu$.
 The function $P_{\rm  H}(E_{\nu},E_{\rm th})$
is thus given by Eq. (\ref{Pe}) and it is essentially 
$P_{\rm H}(E_{\nu},E_{\rm th})\simeq 1$, as we can understood by considering that 
 $E_{\rm G}\gg E_{\rm th}$.

\begin{figure}[t]
\centering
\includegraphics[scale=0.65]{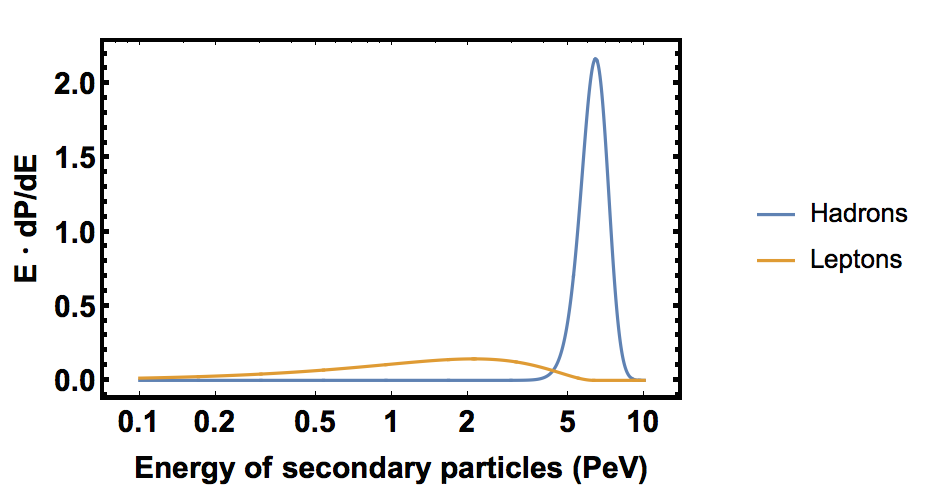}
\caption{\em\small Energy spectra of secondary charged 
particles produced at the Glashow resonance. 
For the hadronic component, the energy resolution of the detector is taken into account. The integrals of fluxes equate the branching ratio of 
Tab.~\ref{brw}, about 2/3 for hadronic channel and about 1/3 for all leptonic channels.}
\label{glash2comp}
\end{figure}

In leptonic decays, a large part of the incoming neutrino energy $E_\nu$
is carried away by the invisible outgoing neutrinos. The charged lepton has a continuous spectrum of energy,
that for any leptonic species is given by:
\begin{equation}
\frac{dP}{dE}=  \frac{3} {E_\nu} \left( 1-\frac{E}{E_\nu} \right)^{2}   \theta(E_\nu - E)
\end{equation}
and it is shown by the yellow line in Fig.\ref{glash2comp}. We see that processes
in which the lepton takes a small fraction of the neutrino energy 
are favored. When $W^-\rightarrow \overline{\nu}_e+e$, 
the electron deposits all its energy into the detector as an
electromagnetic cascade. Neglecting energy resolution effects, 
we evaluate:
\begin{equation}
P_{\nu e}(E_\nu,E_{\rm th}) = \int_{E_{\rm th}} dE \,
(dP/dE) = (1-E_{\rm th}/E_\nu)^3
\label{Pnue}
\end{equation}
When $W^-\rightarrow \overline{\nu}_\tau+\tau$,  the tau deposits
\textcolor{black}{a fraction $x_\tau = 73\%$} of its total energy as electromagnetic and hadronic
cascade; the function $P_{\nu\tau}(E_\nu,E_{\rm th})$ can
be obtained from Eq.(\ref{Pnue}) by replacing \textcolor{black}{$E_{\nu} \rightarrow x_\tau E_\nu $. Note that the factor $x_{\tau}$ is
different from the parameter $\eta_{\nu_\tau}$, defined in Sec.\ref{disec},
that gives the average fraction of incoming neutrino energy in
$\nu_\tau$ CC-interactions which is deposited in the detector. The two
quantities
are related by $\eta_{\nu_\tau} = (1-\langle y \rangle ) x_\tau + \langle y
\rangle$,
where $\langle y \rangle\simeq 1/4$ is the mean inelasticity in $\nu_\tau$
CC-interactions.}

Note that the finite width of the charged lepton energy distributions reduces 
the relative contribution of leptonic modes to cascades produced 
by Glashow resonance above a certain threshold. For $E_{\nu}=E_{\rm G}$ and $E_{\rm th}=2\,{\rm PeV}$, we obtain $P_{\nu e} =0.32$ and
$P_{\nu \tau}=0.18$ showing that, due to threshold effects, the
contribution to the event rate of leptonic modes is reduced by $\sim 75\%$.
By taking into account the branching ratios of the different channels, 
this implies that hadronic modes account for 90 $\%$ of the total signal produced by Glashow resonance. 

\subsection{The effective areas for DIS and for Glashow resonance}
\label{disentangling}

 In order to calculate the number of cascades produced above 2 PeV 
by DIS and Glashow resonance, we need to determine the effective areas
 $A^{\mbox{\tiny DIS}}_e(E_\nu)$,  $A^{\mbox{\tiny DIS}}_\tau(E_\nu)$ and  $A^{\rm G}_e(E_\nu)$ defined in
 Eq. (\ref{DIS}) and (\ref{GR}). 
The simplest way is to consider that, at high energy, 
the DIS cross section is essentially independent on the neutrino flavor. Thus, 
we expect:
\begin{equation}
A^{\mbox{\tiny DIS}}_e(E_\nu)=A^{\mbox{\tiny DIS}}_\tau(E_\nu) =
A_\tau(E_\nu)
\end{equation}
 where we considered that $\nu_\tau$ only interact
through DIS and we implicitly assumed that detection efficiencies
of $\nu_e$ and $\nu_\tau$ are equal above $\sim 1$PeV. 
The effective area for Glashow resonance can then be calculated by
subtraction, obtaining:
\begin{equation}
A^{\rm G}_e(E_\nu) = A_e(E_\nu) - A_\tau(E_\nu).
\end{equation}
Both the total effective areas $A_e(E_\nu)$ and $A_\tau(E_\nu)$ have been calculated by IceCube
and are given in \cite{Aartsen:2013jdh}.
\begin{figure}[t]
\centering
\includegraphics[scale=0.6]{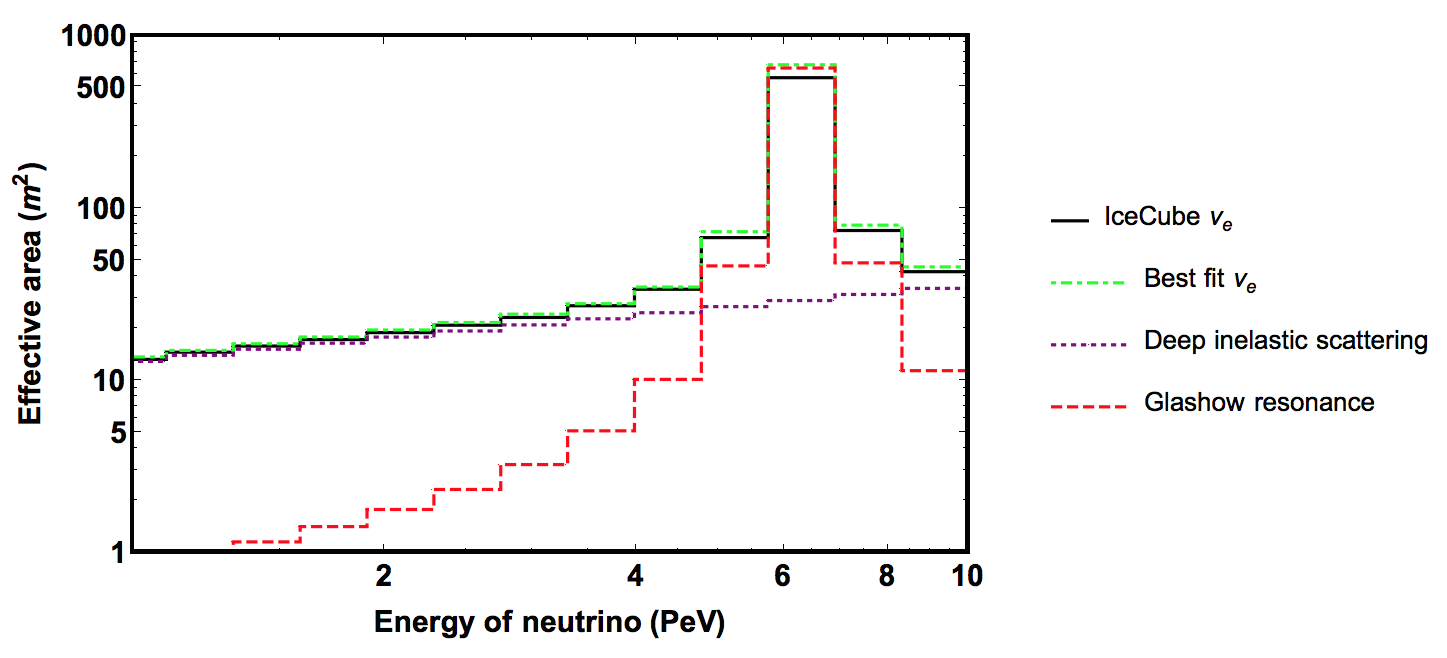}
\caption{\small\em Effective areas of $\nu_e$ at high energies. The effective area given by IceCube (continuous line) is reproduced \textcolor{black}{within 10 $\%$ on average (15 $\%$ in the worst bin)} by the sum of a contribution due to DIS  (dotted line) and the contribution of Glashow resonance (dashed line), discussed in Sect.~\ref{sec3}.}
\label{glashobin}
\end{figure}
\\It is, however, important to understand the main
properties of $A^{\rm DIS}_\ell(E_\nu)$ and
$A^{\rm G}_e(E_\nu)$ on physical basis.  
We expect that:
\begin{equation}
\nonumber
A^{\mbox{\tiny DIS}}_{\ell}(E_\nu) = \epsilon (E_\nu) \left[N_{\mbox{\tiny n}}   \times
                         \sigma_{\mbox{\tiny DIS}}(E_\nu) \times
                         (1+h(E_\nu))/{2}\right] \ \ \ \ \ \mbox{where $\ell=e,\,\tau$}
\end{equation}
where $N_{\mbox{\tiny n}}=\frac{\rho  V}{m_N}=5.5\times 10^{38}$ is
the number of nucleons in $1\,{\rm km}^3$ of ice with density of 0.92 g/cm$^3$,
$\sigma_{\mbox{\tiny DIS}}(E_\nu)=0.89 (E_{\nu}/{\mbox{PeV}})^{0.45}\times$  $10^{-33} \ \mbox{cm}^2$  
is the total DIS (CC+NC) cross section \cite{Gandhi} and \textcolor{black}{we considered negligible the difference between the cross section of $\nu$ and $\bar{\nu}$ that is less than 5 $\%$ for neutrino energy above 1 PeV (see \cite{Gandhi}). Let us remark that both CC and NC cross sections must be included to
reproduce
the IceCube effective areas above 1 PeV, because the effective areas
given in \cite{Aartsen:2013jdh} have been calculated with a low energy
threshold around 30 TeV that does not cut events produced by NC interactions of PeV neutrinos, 
even if the deposited energy is about 1/4 of the incoming
neutrino energy. }
The factor $h(E_\nu)$
describes  neutrino absorption in the Earth,  
modeled using PREM \cite{prem} and averaged over the angle of arrival
of neutrinos\footnote{For energies $E_\nu=$10, 100, 1000 and 10000
  TeV we find that $h=0.91,0.66,0.37$ and $0.18$ respectively.}.
The parameter $\epsilon(E_\nu)$ gives the IceCube effective
volume with respect to an ideal $1\,{\rm km}^3$ detector and includes 
the effects of space, time and energy cuts  in the HESE analysis. 
By comparing the effective areas calculated by IceCube
\cite{Aartsen:2013jdh}, $A_{\tau}(E_\nu)$,
with our estimate, $A^{\mbox{\tiny
  DIS}}_\tau(E_\nu)$, we determine the unknown efficiency $\epsilon(E_\nu)$. The efficiency can be
described as:
\begin{equation}
\epsilon(E_\nu) = 0.40 \ \bigg(\frac{E_\nu}{\mbox{PeV}}\bigg)^{0.075}  \nonumber
\end{equation}
for neutrino energies $1\, {\rm PeV}\le E_\nu \le 10\, {\rm PeV}$. We
see that it is nearly constant, varying by $\sim 15\%$ when $E_\nu$ varies by one decade.
Using this efficiency it is possible to obtain the effective area for Glashow resonance as follow:
\begin{equation}
A^{\rm G}_e(E_\nu) = \epsilon(E_{\nu}) \left[\frac{1}{2} \times 
\frac{1}{2} \times N_{\mbox{\tiny e}}\times \sigma_{\rm G}(E_\nu) \right]
\end{equation}
where $\sigma_{\rm G}(E_\nu)$ is the total $\overline{\nu}_e + e$
cross section and $N_{\mbox{\tiny e}} =10/18 \times N_{\mbox{\tiny    n}}=3.1\times 10^{38}$
is the total number of electrons in $1\,{\rm km}^3$ of ice.
The first factor 1/2 takes into account that only $\overline{\nu}_e$
interact through Glashow resonance and that 
IceCube calculations are obtained by considering an antineutrino
fraction $\tilde{\xi}_{\bar{\nu}_e}=1/2$ \cite{Aartsen:2013jdh}.
The second factor 1/2 is obtained by assuming complete
absorption of antineutrinos crossing the Earth\textcolor{black}{, only for the Glashow resonance piece of the $\nu_e$ effective area.}
In order to verify the adequacy of our interpretation, we compare  in Fig.\ref{glashobin} the IceCube effective area,
$A_e(E_\nu)$, with the sum of the two contributions $A^{\rm G}_e(E_\nu)+A^{\rm
  DIS}_{e}(E_\nu)$. We are able to reproduce $A_e (E_\nu)$ within 10 $\%$ accuracy, showing that the main physical ingredients are correctly understood and implemented. 
The small difference between our parametrization and IceCube calculation near the Glashow resonance could be due to a slightly lower
efficiency of IceCube to detect muons and tau produced by leptonic
channels of the $W^-$ boson.   

\subsection{Results}
\label{risultati}

By using the previous considerations we can obtain the expected number of cascades
above $E_{\rm dep}=2$ PeV as a function of the spectral index $\alpha$ of the
incoming neutrino flux.
The number of events from Glashow resonance is given to a good approximation by the analytical expression:
\begin{equation}
\label{prox2}
N_{\mbox{\tiny G}}(\alpha) \approx 4.75 \times \frac{T}{\mbox{year}} \times  \frac {\xi_{\bar{\nu}_e}}{\tilde{\xi}_{\bar{\nu}_e}} \times F(\alpha) \times \bigg(\frac{E_{G}}{\mbox{PeV}} \bigg)^{2-\alpha}  
\end{equation}
where $T$ is the exposure time and the factor $F(\alpha)$ is the flux
normalization discussed in Eq.(\ref{fa}).
The number of events from DIS can be fitted with the same functional form as:
\begin{equation}
\label{prox3}
N_{\mbox{\tiny DIS}}(\alpha) \approx 4.14 \times \frac{T}{\mbox{years}} \times \bigg(\frac{E_{\mbox{\tiny DIS}}}{\mbox{PeV}} \bigg)^{2-\alpha} \times F(\alpha)
\end{equation}
where the parameter $E_{\mbox{\tiny DIS}}=4.05 \
\mbox{PeV}$. We notice that both Glashow resonance and DIS events depends on $\alpha$ more strongly than double pulse events, as can be seen comparing with Eq.(\ref{prox1}).

The total number of cascades with energy above
2 PeV depends from the production mechanism. 
The relevant parameter is the $\bar{\nu}_e$ fraction
$\xi_{\bar{\nu}_e}$ which determines $N_{\mbox{\tiny G}}(\alpha)$ and
thus fix the relative contribution of events from Glashow resonance and DIS \cite{Anchordoqui:2004eb,Bhattacharya:2011qu,Pakvasa}. 
In the case of $pp$ interactions, about an equal number of neutrinos and
antineutrinos are produced at the source, with flavor ratios
$(1:2:0)$. On the other hand, if the production mechanism
is $p\gamma$ and we consider the simplest scenario, only $\pi^+$ are produced and there are not $\bar{\nu}_e$
at the source. Taking into account neutrinos oscillations
and their uncertainties the fraction of $\bar{\nu}_e$ arriving at Earth
with respect to the total electronic flux is given by:           
$$
\xi_{\bar{\nu}_e}=
\left\{
\begin{array}{ccr}
\displaystyle
\frac{1}{2}+ P_1&= 0.500 \pm 0.029 & \mbox{ if }pp\mbox{ source} \\[2ex]
\displaystyle
\frac{1-3 P_0}{3}+ P_1&= 0.224 \pm 0.029&  \mbox{ if }p\gamma \mbox{ source}
\end{array}
\right.
\label{ciao}
$$
\label{xig}
where $\xi_{\bar{\nu}_e}$ as a function of $P_i$ is obtained in
\cite{flv2}.
This extreme scenario maximises the difference between the signals from $pp$ and $p\gamma$ sources; when we take into account the possibility that some amount of $\bar{\nu}_e$ is also created by $p\gamma$ interactions at the source, the differences diminish. The contamination with $\bar{\nu}_e$ at the source is for $p \gamma$ depending on the target photon spectrum, typically
around 20-50 $\%$ with respect to the flux of $\nu_e$, in the energy range between 1 TeV and 1 PeV (see Fig. 13 of [25] and the paper [26], in which the contaminations of $\bar{\nu}_e$ in $p\gamma$ interactions are discussed in details). In our extreme scenario the two mechanisms give separate
predictions for $\xi_{\bar{\nu}_e}$ even if uncertainties on
oscillation parameters are included.
\begin{figure}[t]
\centering
\includegraphics[scale=0.60]{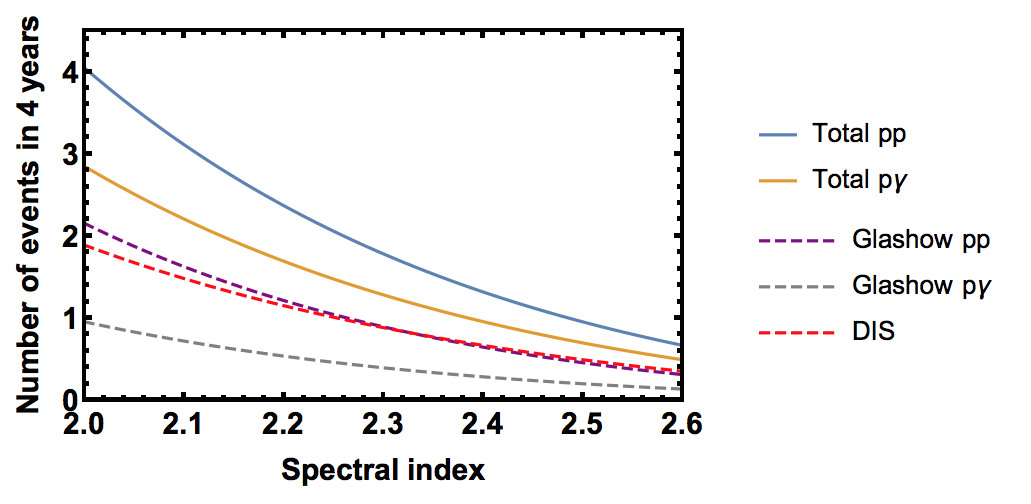}
\caption{\em \small Number of expected cascade events with deposited energy above
  2 PeV with an exposure time of 4 years as a function of the neutrino
  flux spectral index. We show the total number of events for the two
  different production mechanisms. For the Glashow resonance we have chosen the best fit values of $\xi_{\bar{\nu}_e}$, both for $pp$ and $p\gamma$ mechanisms.
}
\label{rateg}
\end{figure}

\begin{figure}[t]
\centering
\includegraphics[scale=0.45]{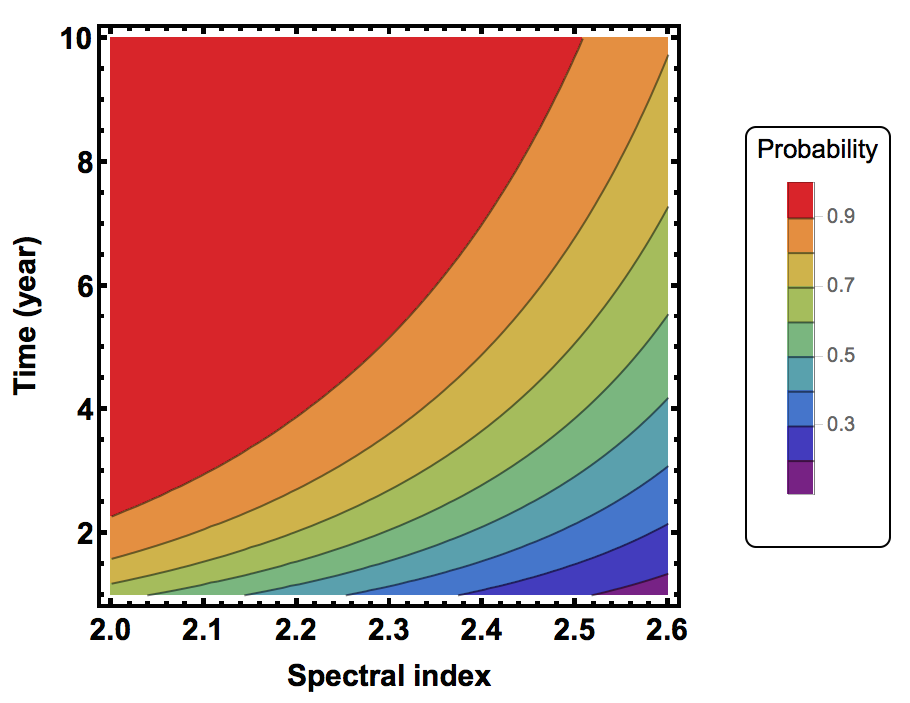}
\includegraphics[scale=0.45]{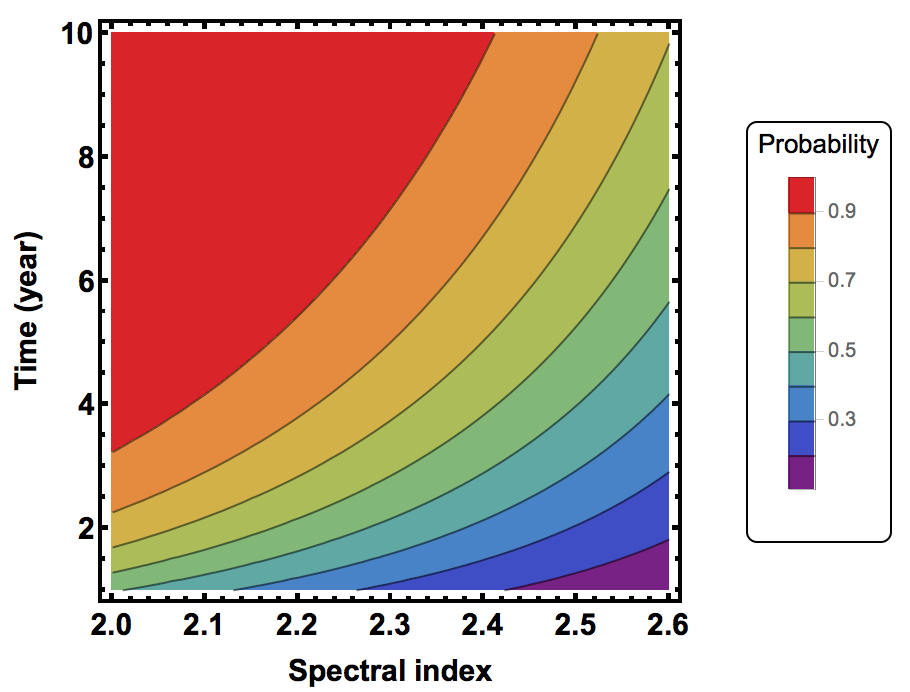}
\caption{\em \small Probability to observe at least one cascade above 2 PeV as a function of spectral index and number of years of exposure. On the left panel $pp$ interaction at the source, on the right panel $p \gamma$ interaction at the source.}
\label{probcasc}
\end{figure}

The total number of cascades above 2 PeV after 4 years is shown in Fig. \ref{rateg}, where we can see the different contributions of DIS and Glashow resonance. 
In the assumption of $pp$ interactions at the source, we obtain 4.0, 2.4, 1.3, 0.7 events expected in 4 years with $\alpha=$2.0, 2.2, 2.4 and 2.6 respectively. These numbers are reduced by $\simeq 25 \%$ for $p \gamma$ interactions. In Fig.\ref{probcasc} we show the probability to observe at least one event as a function of the assumed spectral index and of the observation time. The non observation of cascades above 2 PeV is in tension with the hypothesis of an hard neutrino spectrum. An an example, for $pp$ mechanism, $\alpha < 2.2$ is excluded at 90 $\%$ CL. 

It is evident from the above results that, unless the neutrino spectral index is fixed, we cannot discriminate between different neutrino production mechanisms. In fact, the indetermination of the event rate due to our incomplete knowledge of the neutrino spectrum, is comparable with the differences generated by the various production mechanisms. 
It is interesting to note that, also for a fixed value of
the spectral index, the number of expected events is so small that an exposure of tens of years is required. For example, focusing on events around
6.32 PeV, where the background due to DIS is negligible, and with
$\alpha=2.3$, we need more than 50 years to
obtain some constraint a 2$\sigma$ discrimination between $pp$ and $p \gamma$ interaction at the source. Of course this time increases if some $\bar{\nu}_e$ are produced at the source also by the $p\gamma$ mechanism, reducing the differences between $pp$ and $p\gamma$ interaction.
For these reasons, we think that nowadays every conclusion about the
mechanism of production is just a speculation and only a detector with
a bigger exposure can clarify the situation in the future. 


Previous results are obtained by assuming an unbroken power law for
the neutrino flux in the energy region below 10 PeV. The presence of
an energy cutoff below $E_{\rm G}$ drastically decreases the number of
events due to the Glashow resonance, whereas reduces the DIS events of
only about 20 $\%$.
In the opposite case in which the energy cutoff is much greater than
10 PeV the number of Glashow events is not affected, whereas the
number of DIS events increases of about 30-40 $\%$. 

\textcolor{black}{To determine our ability to distinguish between an unbroken power law with $\alpha=2.3$ from a power law with cutoff below $E_{\rm G}$ we used Poissonian statistics $P(n,\mu)=\frac{e^{-\mu} \mu^n}{n !}$. The absence of events due to the Glashow resonance around 6.32 PeV with an energy resolution of 10 $\%$ in the deposited energy, is to date compatible with both the spectral shapes. To have an hint at $2\sigma$ of a cutoff, we need to wait 15 years for $pp$ interaction at the source. Indeed this corresponds to $\mu=3$ expected events, that means a probability $P(0,3)=e^{-3}$ less than $5\%$ to see no events. Of course the required number of years is less than 15 years for $\alpha < 2.3$ and greater for $\alpha > 2.3$ and it doubles in the case of $p\gamma$ mechanism of production.}


\section{Discussion}
\label{sec4}

To summarize we report in the Tab. \ref{sommario} the number of
IceCube expected events for 1 year of exposure and for the different
channels discussed in this work. Moreover we report in the last two
columns the ratios between the number of cascades above 2 PeV and
double pulse events.     
\begin{table}[b]
\caption{\em \small Expected number of events per year for the different classes
  of events by using the best fit value of $\xi_{\bar{\nu}_e}$ and $(\nu_e:\nu_\mu:\nu_\tau)=(1/3:1/3:1/3)$ at
the Earth. \textcolor{black}{$N_{\mbox{\tiny DIS}}$ and $N_{\mbox{\tiny G}}$ refer to showers with deposited energy above 2 PeV, $N_{\mbox{\tiny 2P}}$ refers to double pulses without threshold. }}
\label{sommario}
\begin{center}
\begin{tabular}{|c|c|c|c|c|c|c|}
\hline
$\alpha$ & $N_{\mbox{\tiny 2P}}$ & $N_{\mbox{\tiny DIS}} $ & $N_{\mbox{\tiny
G}}^{\mbox{\tiny pp}}$ & $N_{\mbox{\tiny G}}^{\mbox{\tiny p$\gamma$}}$ &
($N_{\mbox{\tiny G}}^{\mbox{\tiny pp}}+N_{\mbox{\tiny DIS}})/N_{\mbox{\tiny 2P}}$ & $(N_{\mbox{\tiny
G}}^{\mbox{\tiny p$\gamma$}}+N_{\mbox{\tiny DIS}})/N_{\mbox{\tiny 2P}}$ \\
\hline 
2 & 0.165 & 0.473 & 0.539 & 0.239 &6.116  & 4.306 \\
\hline
2.2 & 0.132 & 0.288 & 0.304 & 0.135 & 4.504 & 3.220 \\
\hline
2.4 & 0.103 & 0.168 & 0.162 & 0.072 &3.192 & 2.319 \\
\hline
2.6 & 0.078 & 0.089 & 0.079 & 0.035 &2.182 & 1.613 \\
\hline
\end{tabular}
\end{center}
\label{default}
\end{table}%
The ratios of the last two columns decreases with $\alpha$, because Glashow resonance and deep inelastic scattering are more affected by a change of the slope with respect to the number of
double pulse events, as we can see in the Eqs.(\ref{prox1},\ref{prox2},\ref{prox3}). 

\begin{figure}[t]
\centering
\includegraphics[scale=0.7]{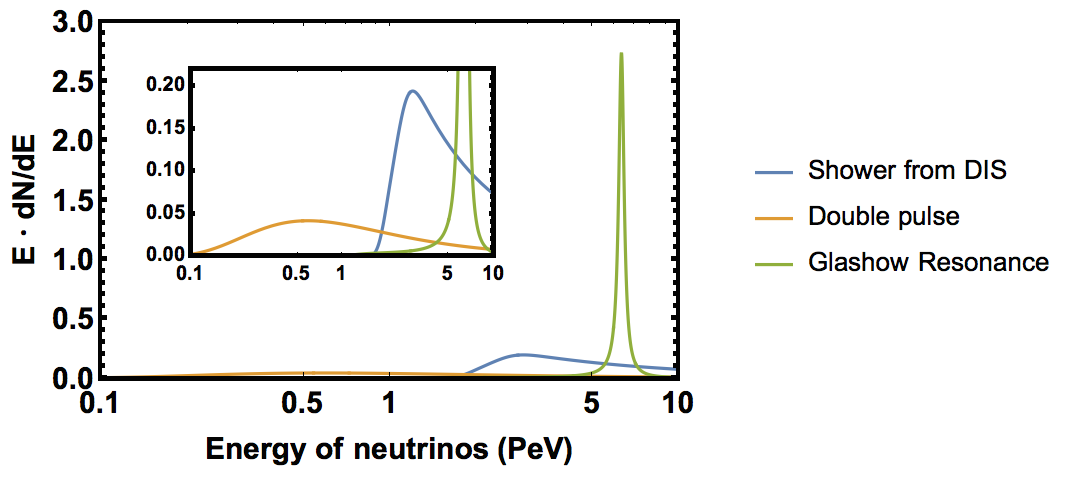}
\caption{\em \small $E \frac{dN}{dE}$ for the three types of processes analyzed in the paper, under the hypothesis $\alpha=2.3$ in the neutrino spectra.}
\label{parental}
\end{figure}

In order to clarify what is the relevant energy region of the parental
neutrino flux giving a contribution to expected events discussed in
this work, we show in Fig.\ref{parental} the integrands $dN/dE_{\nu}$
of Eqs.(\ref{tau1}-\ref{DIS}-\ref{GR}) multiplied by an extra factor $E_\nu$ for
the three channels analyzed in this paper for a spectral index
$\alpha=2.3$. 
It is interesting to remark that also $\nu_\tau$ of about 0.5 PeV can
give rise to a "double pulse" event, because there is a compensation 
between the flux power law decrease and the probability to not decay 
that increases exponentially with the energy. 

The difference between $pp$ and $p \gamma$ interaction at the source is already discussed in other works. We compare here with some of them, that are important and well-known, namely
 \cite{Pakvasa}, \cite{Anchordoqui:2004eb}, \cite{Bhattacharya:2012fh} and \cite{Barger:2012mz}.
We point out the reasons why our analysis is different and show that, although results could seem to be incompatible, there are no contradictions.
Our analysis of the events with deposited energy above 2 PeV
shows that the rate in the case of $p \gamma$ interaction at the source is about 75 $\%$ of the rate given by $pp$ interaction. In fact, the difference is due to events produced by  Glashow resonance, but 
in this region of energy, the rate of DIS events is large and this reduces the difference between the two types of mechanism. 
This is not in contradiction with what is written in the paper \cite{Pakvasa}, in which the ratio between resonant and non resonant event above 2 PeV, is about one half. In that analysis, in fact,
{\em the energy of incoming neutrino and not the deposited energy into the detector} is considered. When the deposited energy is considered instead, the neutral current interaction of all type of neutrinos and the charged current interaction of $\nu_\mu$ give a negligible contribution to cascade above 2 PeV: this explain the difference between the results of the two papers.  Our analysis uses 
the deposited energy, since this is observable, and
it is not simple (or possible) to reconstruct the energy of incoming neutrinos. That explain also the difference with respect to \cite{Guetta}.

The comparison with  \cite{Anchordoqui:2004eb} concerns
another important remark. Even in this work, the difference between $pp$ and $p \gamma$ interaction at the source is discussed and their conclusion is that the number of Glashow events, in the case of $pp$ mechanism, is about 6 times greater than what expected in the case of $p \gamma$ mechanism. Again, this statement is not in contradiction with our statement, because {\em they compare 
models with the same flux of protons at the source} whereas we take into account the observed flux of neutrinos at the Earth. In other words, we ask ourselves what is the fraction of electron antineutrinos with respect to the {\em observed} total flux of neutrinos and, as a consequence, only the difference due to neutrinos oscillations is relevant for these considerations. The oscillations produce a difference of about a factor of 2 between the two mechanisms of production, and the additional difference found by  
\cite{Anchordoqui:2004eb} is due to the different fraction of energy 
that neutrinos receive in the $pp$ and $p \gamma$ production mechanism.

The last remark concerns a hypothesis proposed in  \cite{Bhattacharya:2012fh} and \cite{Barger:2012mz}, where it was suggested that the events observed above PeV could be due to Glashow resonance. 
If this was true, were caused by leptonic decay, since the energy resolution is incompatible with a visible energy being 6.32 PeV.
Now, the 
ratio between hadronic and leptonic branching ratios is,
${\Gamma(hadr)}/3 \  \Gamma(l \ \nu)\simeq 2$
and the same neutrino flux leads to leptonic and hadronic decays. 
To date there are three events with energy below 2 PeV, so in the above assumption, we expect to have also 6 hadronic events. 
The probability to observe none is given by the Poissonian PDF, $P(0)=e^{-6} $, which is  disfavored at 3 $\sigma$.


\section{Conclusion}
\label{sec5}
 In this paper we have considered  two categories of very high energy
 events in IceCube that can help to diagnose cosmic neutrinos:
the double pulse events, that may allow us to clearly discriminate the
cosmic component of $\nu_\tau$;
the cascades with deposited energy above 2 PeV,
including events produced by $\overline{\nu}_e$ at the Glashow resonance,
that can be used
to investigate the cosmic neutrino production mechanisms.

 As stated in the Introduction, we estimated the rate of these high energy
events
with the important constraint provided by the data already observed by
IceCube,
i.e. we used the data collected in the low-energy region below $2$PeV
to normalize our calculations. In this way, we obtained the expected
rates of high energy events as a function of the neutrino spectral
index $\alpha$, that we varied in the range $\alpha=(2-2.6)$.

 We found that the non-observation of double pulse events does not
contradict the hyphotesis of a cosmic neutrino population. This conclusion
is only marginally dependent on the
assumed cosmic neutrino spectrum. In fact, we have shown that:\\
\textit{i)}  One half
of the expected signal is due to neutrinos with energy below $E_\nu=2$
PeV, i.e.
\textit{from a spectral region that is already observed in the HESE
  data} (see Fig. \ref{parental} and discussion in the section
\ref{sec2})\\
\textit{ii)}
In the most favorable case, with spectral index $\alpha=2$ \textit{we need
to wait about 10 more years to observe a double pulse
with a probability greater than 90 $\%$}.

 Concerning the cascades with deposited energy above 2 PeV,
we have shown that:\\
\textit{i)} Due to the difference between the energy deposited in the
detector and the energy of the interacting neutrino, the contribution of
leptonic channels to Glashow resonance events is suppressed by
75$\%$. This implies that hadronic modes account for 90$\%$
of the total signal produced by Glashow resonance above 2 PeV;\\
 \textit{ii)} This class of events can be used to probe the high energy
tail of the cosmic neutrino
spectrum. The absence of cascades above 2 PeV disfavours a neutrino
spectral index $\alpha=2$ (with a cutoff $E_{\rm cut}\ge 10$PeV)
at about 2$\sigma$ considering that 3 events are expected in the worst
scenario ($p \gamma$ interaction).
An unbroken power law with $\alpha=2.6$ is instead still compatible at
1$\sigma$ with the present results. The absence of events close to the
Glashow resonance energy $E_{\rm G}= 6.32$PeV is not problematic under
this hypothesis, since only
0.4 events are expected due to Glashow resonance after 4 years (see the
table \ref{sommario}); \\
\textit{iii)} the difference between the event rates produced by $pp$
or $p\gamma$ neutrino production mechanisms is not large enough to
distinguish among the two options, even if we assume that the parent
neutrino
spectrum is known. An observation time $T\sim 50$ years would be
required to obtain a $2\sigma$ discrimination, if the neutrino
spectral index is $\alpha=2.3$.

\end{document}